\documentclass[10pt,letterpaper,twocolumn]{article} 

\usepackage{ol2}
\usepackage[draft]{hyperref}
\usepackage{amsmath}

\begin{document}

\twocolumn[ 

\title{Hybrid surface waves in semi-infinite metal-dielectric lattices}

\author{Juan J. Miret,$^1$ Carlos J. Zapata-Rodr\'{\i}guez,$^{2,*}$ \\ Zoran Jak\u{s}i\'c,$^3$ Slobodan Vukovi\'c,$^{3,4}$ and Milivoj R. Beli\'c$^4$}

\address{
$^1$Department of Optics, Pharmacology and Anatomy, University of Alicante, Spain \\
$^2$Department of Optics, University of Valencia, Dr. Moliner 50, 46100 Burjassot, Spain \\
$^3$Center of Microelectronic Technologies and Single Crystals, Institute of Chemistry, Technology and Metallurgy, \\
University of Belgrade, Njego\u{s}eva 12, 11000 Belgrade, Serbia \\
$^4$Texas A \& M University at Qatar, P.O. Box 23874, Doha, Qatar \\
$^*$Corresponding author: carlos.zapata@uv.es
}

\begin{abstract}
We investigate surface waves at the boundary between a semi-infinite layered metal-dielectric nanostructure cut normally to the layers and a semi-infinite dielectric.
Spatial dispersion properties of such a nanostructure can be dramatically affected by coupling of surface plasmons polaritons at different metal-dielectric interfaces.
As a consequence, the effective medium approach is not applicable in general.
It is demonstrated that Dyakonov-like surface waves with hybrid polarization can propagate in an angular range substantially enlarged compared to conventional birefringent materials.
Our numerical simulations for an Ag-GaAs stack in contact with glass show a low to moderate influence of losses
\end{abstract}


 ] 

\noindent Currently, metallodielectric superlattices are receiving increased attention because of their extraordinary optical properties such as near-field focusing, subwavelength imaging, and negative refraction \cite{Fan06,Ceglia08,Pastuszczak11}.
The inclusion of metallic elements is responsible for the excitation of plasmonic resonances on the metal-dielectric (MD) interfaces, giving rise to an anisotropic metamaterial.
Surface polaritons may exist also along its optical axis, when the plasmonic crystal is cut normally to the orientation of the layers \cite{Vukovic09}.

Hybrid-polarized waves confined to the boundary of a semi-infinite layered periodic nanostructure, and importantly with oblique propagation with respect to the lattice optical axis, have first been reported in the context of hyperlensing \cite{Jacob08}.
The effective-medium approximation \cite{Rytov56} (EMA) was applied to benefit from a simple approach that leads to Dyakonov surface waves \cite{Dyakonov88}.
In this letter we demonstrate that in general, the use of the EMA is not justified.
Specifically, we focus on surface waves at the boundary between a MD nanostructure and a semi-infinite dielectric.
By solving Maxwell's equations, we demonstrate the existence of Dyakonov-like surface waves with hybrid polarization and establish that they can propagate in a much wider angular range than the conventional birefringent materials.
In our calculations, realistic metal widths and dissipation effects are utilized.

\begin{figure}[t]
 \centerline{\includegraphics[width=6cm]{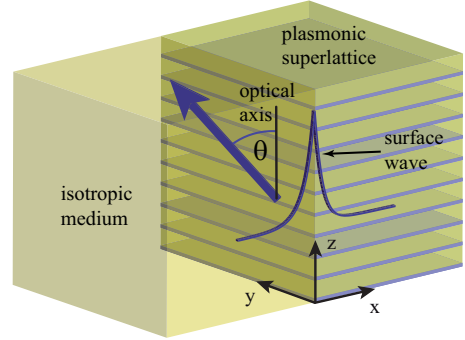}}
 \caption{Schematic setup under study, consisting of a semi-infinite Ag-GaAs superlattice ($x > 0$) 
and an isotropic cover ($x < 0$), either N-BAK1 or P-SF68 [SCHOTT].}
 \label{fig01}
\end{figure}

We consider a bilayered superlattice made of two materials alternatively stacked along the $z$ axis (Fig.~\ref{fig01}).
The unit cell of the 1D photonic lattice consists of a transparent material of dielectric constant $\epsilon_d$ and slab width $w_d$, followed by a metallic layer with the corresponding parameters $\epsilon_m$ and $w_m$.
In our numerical simulations we set $\epsilon_d = 12.5$ and $\epsilon_m = -103.3 + i 8.1$, which correspond to GaAs and Ag, respectively, at the wavelength $\lambda_0 = 1.55\ \mu\mathrm{m}$.
This metamaterial fills the semi-space $x > 0$. Adjacent to the periodic medium, filling the space $x < 0$,
is an isotropic material of dielectric constant $\epsilon$. 
At the boundary between the periodic and the isotropic medium, in the plane $x = 0$, we expect to find bounded waves with evanescent amplitudes as $|x| \to \infty$.

We seek solutions of Maxwell's equations in the form of surface waves that can propagate at the MD-superlattice boundary.
First, we configure the optical anisotropy of our periodic structure by employing average estimates.
In particular, the form of birefringence in this type of plasmonic devices may be modeled by the EMA \cite{Yeh88}.
The validity of this approximation is related to the fact that layer widths are much shorter than the wavelength, $w_d,w_m \ll \lambda_0$.
In this case, the superlattice behaves as a uniaxial crystal whose optical axis is normal to the layers (here, $z$-axis).
The model estimates the relative permittivities along the optical axis, $\epsilon_{||} = \epsilon_m \epsilon_d / \left[ \left( 1 - f \right) \epsilon_m + f \epsilon_d \right]$,
and transversally, $\epsilon_\perp = \left( 1 - f \right) \epsilon_d + f \epsilon_m$,
where $f$ is the metal filling factor, $f = w_m / \left( w_d + w_m \right)$.
If we neglect losses by setting $\mathrm{Im}(\epsilon_m) = 0$, the effective birefringence of the Ag-GaAs superlattice is $\Delta n = \sqrt{\epsilon_{||}} - \sqrt{\epsilon_\perp}$.
Even small filling factors of the metallic composite lead to enormous birefringences.
If $f \ll 1$ and $|\epsilon_m| \gg \epsilon_d$, we obtain $\Delta \epsilon \approx - \epsilon_m f$, which holds up to $f_\mathrm{max} = 0.108$, where $\epsilon_\perp = 0$.

Filling factors higher than $f_\mathrm{max}$ provide negative values of $\epsilon_\perp$ and, as a consequence, effective dispersion becomes hyperbolic \cite{Wood06}.
Thus $\Delta n > 0$, since $\epsilon_m < 0$.
In spite of $f_\mathrm{max} \ll 1$, the maximum birefringence achievable reaches $\left[ \Delta n \right]_\mathrm{max} = 3.77$.
For comparison, we quote the birefringence $\Delta n = 0.0084$ for crystalline quartz and $\Delta n = 0.22$ for liquid crystal BDH-E7 \cite{Wu84}.
Thus, our artificial uniaxial crystal has birefringence greater by more than an order of magnitude.

Since we treat the 1D lattice as a uniaxial crystal, we may establish the diffraction equation giving the wave vector of the surface wave at $x = 0$.
For that purpose we follow Dyakonov \cite{Dyakonov88}, by considering hybrid surface modes.
In the isotropic medium, we consider homogeneous TE and TM waves in the plane $x = 0$, whose wave vector has real components $k_y$ and $k_z$.
However, these fields are evanescent in the isotropic medium, proportional to $\exp \left( - \kappa |x| \right)$, where $\kappa = \sqrt{k_y^2 +k_z^2 - k_0^2 \epsilon}$ and $k_0 = 2 \pi / \lambda_0$.
On the other side of the boundary, the ordinary and extraordinary waves in the \emph{effective} uniaxial medium also exponentially decay, with rates given by
$\kappa_o = \sqrt{k_y^2 +k_z^2 - k_0^2 \epsilon_\perp}$ and $\kappa_e = \sqrt{k_y^2 + k_z^2 \epsilon_{||} / \epsilon_\perp - k_0^2 \epsilon_{||}}$, respectively.
The Dyakonov equation,
\begin{equation}
 \left( \kappa + \kappa_e \right) \left( \kappa + \kappa_o \right) \left( \epsilon \kappa_o + \epsilon_\perp \kappa_e \right)
  = \left( \epsilon_{||} - \epsilon \right) \left( \epsilon - \epsilon_\perp \right) k_0^2 \kappa_o ,
\label{eq01}
\end{equation}
providing a map of the allowed $(k_y,k_z)$ values, has been derived by applying the standard electromagnetic field boundary conditions at $x = 0$.
Assuming that $\epsilon_{||}$, $\epsilon_\perp$ and that all decay rates are positive, the following additional restriction can be deduced
\begin{equation}
 \epsilon_\perp < \epsilon < \epsilon_{||} ,
\label{eq03}
\end{equation}
for the existence of surface waves.
As a consequence, positive birefringence is mandatory to ensure the existence of surface waves.
Therefore, such a layered superlattice featuring surface waves cannot be formed by all-dielectric materials.

To illustrate the difference between using conventional birefringent materials and plasmonic crystals, we solve Eq.~(\ref{eq01}) for a liquid crystal E7 with $\epsilon_{||} = 2.98$ and $\epsilon_\perp = 2.31$ at a wavelength of $\lambda_0 = 1.55\ \mu\mathrm{m}$, and an N-BAK1 cover of dielectric constant $\epsilon = 2.42$.
In this case, Dyakonov surface waves propagate in a narrow angular region $\Delta \theta = \theta_\mathrm{max} - \theta_\mathrm{min}$, where $\theta$ stands for the angle between the vector $(k_y,k_z)$ and the optical axis.
Specifically, we estimate $\Delta \theta = 0.92^\circ$ around a mean angle $\bar \theta = 26.6^\circ$.
Apparently, the angular range $\Delta \theta$ is small; however it is extremely high if compared with that obtained for other optical crystals, like quartz, exhibiting standard birefringence.
To increase the angular spread $\Delta \theta$ even more, we consider Ag-GaAs crystal that has values of $\epsilon_{||} = 14.08$ and $\epsilon_\perp = 0.92$, with a filling factor $f = 0.10$; the birefringence now yields $\Delta n = 2.79$.
In this case, Eq.~(\ref{eq01}) provides solutions in the region of angles from $\theta_\mathrm{min} = 39.0^\circ$ up to $\theta_\mathrm{max} = 71.3^\circ$.
As a consequence, the angular range $\Delta \theta = 32.3^\circ$ has increased by more than an order of magnitude.

\begin{figure}[t]
 \centerline{\includegraphics[width=8cm]{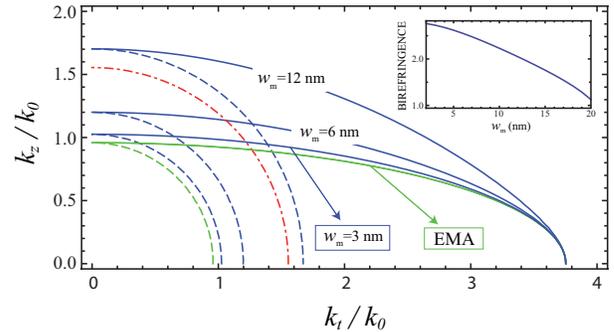}}
 \caption{Isofrequency curves evaluated from Eq.~(\ref{eq05}) for Ag-GaAs superlattice with filling factor $f = 0.10$, 
 for different $w_m$; $k_t = \sqrt{k_x^2 + k_y^2}$.
 Solid lines and dashed lines correspond to TM and TE polarization.
 The dashed-dotted line represents the isofrequency curve of isotropic N-BAK1 cover.
 Inset: Birefringence as a function of $w_m$.}
 \label{fig04}
\end{figure}

The EMA is limited to metallic slabs with $w_m \ll \lambda_0$.
However, this condition must be taken into account with care, since the skin depth of noble metals is extremely short, $\lambda_s \approx c / \omega_p$, where $\omega_p$ is the plasma frequency of the metal.
In the case of silver, we estimate $\lambda_s = 24\ \mathrm{nm}$.
If the thickness of metallic layer is comparable to the skin depth, the EMA will substantially deviate from exact calculations.
Note that experimental multilayers rarely incorporate metallic slabs with a thickness below $10\ \mathrm{nm}$.
To address the problem more thoroughly, we consider the full-wave solution of Maxwell's equations, leading to the spectral map of Bloch waves in bulk 1D-periodic media.
In this case $k_z$ represents the pseudomomentum of a Bloch wave.
In Fig.~\ref{fig04} we plot the spatial dispersion relation \cite{Yeh88}
\begin{equation}
 \cos (k_z \Lambda) = \cos \varphi_m  \cos \varphi_d - \eta_{o,e} \sin \varphi_m  \sin \varphi_d ,
\label{eq05}
\end{equation}
corresponding to TE-polarized (ordinary) and TM-polarized (extraordinary) waves propagating within the periodic Ag-GaAs structure displayed in Fig.~\ref{fig01}.
In Eq.~(\ref{eq05}) the period is $\Lambda = w_m + w_d$, while
$\eta_o =  \left( k_{dz}^2 + k_{mz}^2 \right) / 2 k_{dz} k_{mz} $ and
$\eta_e =  \left( \epsilon_m^2 k_{dz}^2 + \epsilon_d^2 k_{mz}^2 \right) / 2 \epsilon_m \epsilon_d k_{dz} k_{mz}$,
are the coefficients applied for ordinary and extraordinary waves, respectively.
Finally, $\varphi_q = k_{qz} w_q$, where $k_x^2 + k_y^2 + k_{qz}^2= \epsilon_q k_0^2$ represents the dispersion equation for bulk waves within GaAs ($q = d$) and silver ($q = m$).

In the case of negligible losses, as illustrated in Fig.~\ref{fig04}, we observe that EMA is sufficiently accurate for $w_m = 3\ \mathrm{nm}$.
However, deviations among the contours are evident for higher widths.
Apparently, Eq.~(\ref{eq05}) is in good agreement with the EMA in the vicinity of $k_z = 0$ for TM waves only.
In contrast, propagation along the $z$-axis, where $k_x = k_y = 0$, results in larger discrepancies, as the Bloch wave number $k_z$ increases for higher $w_m$.
This effect is observed simultaneously for TM and TE waves.
Consequently, the size of birefringence displayed by TM waves is reduced (see inset in Fig.~\ref{fig04}).
For TE waves, the isotropy of the isofrequency curve is practically conserved, e.g. $n_\perp = 1.70$ and $n_{||} = 1.67$ for $w_m = 12\ \mathrm{nm}$.
At the same time, $\epsilon_\perp$ increases with $w_m$.

Moderate changes in birefringence can substantially impact the existence of Dyakonov-like surface waves.
Enlargement of $\epsilon_\perp$ by increasing $w_m$ (with the same $f$) leads to a significant modification of the isofrequency curve derived from Eq.~(\ref{eq01}).
According to Eq.~(\ref{eq03}), this may entirely prevent the existence of surface waves.

Up to now we have avoided another important aspect of plasmonic devices, that of dissipation in metallic elements.
In order to tackle this problem, we evaluate numerically the value of $k_z$ for a given $k_y$.
Since the imaginary part of $\epsilon_m$ is not neglected anymore, $k_z$ becomes complex.
This means that the surface wave cannot propagate indefinitely; there exists an energy attenuation length $l = (2 \mathrm{Im}[k_z])^{-1}$.
Furthermore, we assume that the real parts of the parameters $\kappa$, $\kappa_o$ and $\kappa_e$ are all positive, which correlates with the field decay at $|x| \to \infty$ and thus with the confinement near $x = 0$.

\begin{figure}[t]
 \centerline{\includegraphics[width=8cm]{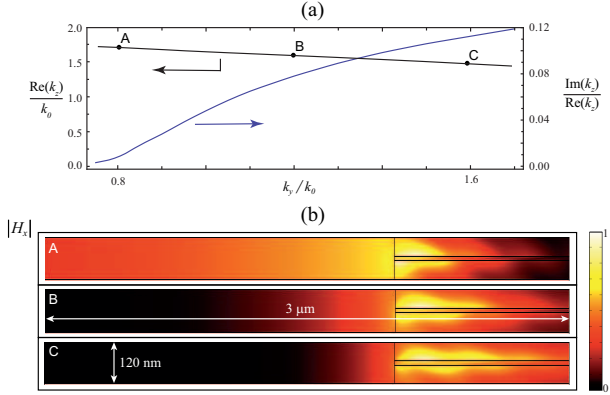}}
 \caption{(a) Isofrequency curve corresponding to hybrid surface waves existing at the boundary between a semi-infinite P-SF68 cover and a plasmonic Ag-GaAs superlattice of $f = 0.10$ and $w_m = 12\ \mathrm{nm}$.
 (b) Three contour plots of the magnetic field $|H_x|$ computed by COMSOL.}
 \label{fig05}
\end{figure}

Fig.~\ref{fig05}(a) shows an isofrequency curve corresponding to Dyakonov-like solutions, for the Ag-GaAs lattice with $f = 0.10$ and $w_m = 12\ \mathrm{nm}$.
The numerical simulations were performed using COMSOL Multiphysics software based on the finite-element method.
In our simulations we did not observe surface waves for a N-BAK1 cover with $n = 1.56$, since now it is not consistent with Eq.~(\ref{eq03}); in our estimations $n_\perp = 1.70$ and $n_{||} = 3.75$.
In particular, Fig.~\ref{fig05}(a) depicts the isofrequency curves when $n = 1.95$, corresponding to P-SF68 [SCHOTT].
The boundaries of such isofrequency curve are established according to the propensity of electromagnetic field to be confined in the neighborhood of $x = 0$.
Fig.~\ref{fig05}(a) also shows $\mathrm{Im}(k_z)/\mathrm{Re}(k_z)$ in the range of existence of the surface waves.
For paraxial surface waves, in which $k_y$ reaches the minimum (case A), we observe that $\mathrm{Im}(k_z) \ll \mathrm{Re}(k_z)$.
This is caused by a large shift of the field maximum toward the isotropic medium, as shown in Fig.~\ref{fig05}(b), which is consistent with $\mathrm{Re}(\kappa) \ll \mathrm{Im}(\kappa)$.
On the other hand, for nonparaxial waves, having the largest values of $k_y$ (case C), the fields show slow energy decay inside the plasmonic superlattice.
As a consequence, losses in the metal become manifested by a significant rise in the values of $\mathrm{Im}(k_z)$.

We conclude that oblique surface waves may propagate at the boundary between a plasmonic bilayer superlattice and an isotropic transparent material.
These modes are not TM-polarized, since all three spatial components of the electric, as well as of the magnetic field are involved, i.e. the modes are hybrid.
Realistic slab widths lead to solutions that deviate significantly from the results of EMA and Dyakonov analysis.
Our numerical simulations prescribe the use of cover materials of a higher refractive index, for the existence of surfaces waves.
A wide angular range of surface waves is attainable with large to moderate energy attenuation lengths.
Finally, let us point out that the properties of the resulting bound states change rapidly with the refractive index of the surrounding medium (cover), which suggests potential applications for chemical and biological sensors.

This research was funded by the Spanish Ministry of Economy and Competitiveness under the project TEC2009-11635, by the Qatar National Research Fund  under the project NPRP 09-462-1-074, and by the Serbian Ministry of Education and Science under the projects III 45016 and TR 32008.


\pagebreak

\section*{Informational Fourth Page}
In this section, we provide full versions of citations to assist reviewers and editors.


\end{document}